# Large Anisotropic Thermal Expansion Anomaly near the Superconducting Transition Temperature in MgB$_2$

J. D. Jorgensen, D. G. Hinks, Materials Science Division, Argonne National Laboratory, Argonne, IL 60439

P. G. Radaelli, W. I. F. David, R. M. Ibberson, ISIS Facility, Rutherford Appleton Laboratory, Chilton, Didcot, Oxfordshire, OX11 0QX, United Kingdom

ABSTRACT

An anisotropic lattice anomaly near the superconducting transition temperature, $T_c$, was observed in MgB$_2$ by high-resolution neutron powder diffraction. The *a*-axis thermal expansion becomes negative near $T_c$, while the *c*-axis thermal expansion is unaffected. This is qualitatively consistent with a depletion of the boron-boron σ-band as the superconducting gap opens, resulting in weaker bonding. However, the observed anomaly is much larger than predicted by the Ehrenfest relation, strongly suggesting that the phonon thermal expansion also changes sign, as commonly observed in hexagonal layered crystals. These two effects may be connected through subtle changes in the phonon spectrum at $T_c$.

INTRODUCTION

Immediately after the discovery of superconductivity in $MgB_2$ [1], we measured its thermal expansion using neutron powder diffraction.[2]  Between the superconducting transition temperature, $T_c$, and room temperature, the thermal expansion can be adequately modeled with an Einstein equation using a single phonon energy.  The thermal expansion along the *c* axis is about twice that along the *a* axis, as a result of the differences in bonding strengths in these directions in the layered hexagonal crystal structure.  Beginning perhaps a couple degrees above $T_c$, we observed a negative change in the volume thermal expansion.  Our measurements of the *a*- and *c*-axis lattice parameters versus temperature suggested that this change resulted almost exclusively from a thermal expansion anomaly along the *a* axis (i.e., in the basal plane of the hexagonal diboride structure).  Unfortunately, the precision of our measurements was not good enough for a careful comparison with theory.

A change in thermal expansion at $T_c$ is expected and its magnitude can be predicted by thermodynamics.  The crystal structure of a superconductor adjusts itself, below $T_c$, in order to minimize the total free energy of the lattice including the superconducting electrons.  Clearly, in the case of an aniostropic superconductor, the change in thermal expansion will reflect the anisotropy.  The expected change in volume thermal expansion at $T_c$ is given by the Ehrenfest relation for a second-order phase transition [3]:

$$\left(\frac{\partial V_s}{\partial T}\right)_P - \left(\frac{\partial V_n}{\partial T}\right)_P = \frac{\partial T_c}{\partial P} \frac{(C_s - C_n)}{T_c} \tag{1}$$

where $V_s$ and $V_n$ are the unit cell volume in the superconducting and normal states, $(C_s-C_n)$ is the change in specific heat at $T_c$, and P is the pressure.

Millis and Rabe [4] showed that fluctuations near $T_c$ could add additional components to the anomaly in thermal expansion.  Fluctuations can explain how one might observe a change in thermal expansion beginning at a temperature above $T_c$, plus enhanced effects below $T_c$.  Fluctuation effects are expected to be important in superconductors where the correlation length is short, such as the layered copper oxides.

These expected effects manifest in the thermal expansion near $T_c$ have been observed and compared with theory for both conventional and exotic superconductors.[5,6]  As expected, the Ehrenfest relation is obeyed in conventional superconductors [5] and additional effects, beginning at temperatures above $T_c$, are seen in the layered copper-oxide superconductors.[6]  For example, in the $YBa_2Cu_3O_7$ superconductor, fluctuation effects, added to the effect predicted by the Ehrenfest relation, are manifest in the thermal expansion beginning about 5 K above $T_c$ and extending about 5 K below $T_c$.[6]  However, even when fluctuation effects increase the size of the change in thermal expansion at $T_c$, the effects are very small.  In general, dilatometry measurements on single crystal samples, which provide a level of sensitivity well beyond that achieved by diffraction measurements, are required to observe these effects and compare them with theoretical predictions.  Thus, our observation of a change in the thermal expansion of $MgB_2$ near $T_c$ in our diffraction measurement [2] was somewhat unexpected.  In this paper, we report additional neutron powder diffraction data, taken on an instrument of significantly better resolution.  Our new results confirm the previous observation of an unexpectedly large, and very anisotropic, change in the thermal expansion of $MgB_2$ very near $T_c$.  The change in volume thermal expansion is about five times larger than predicted by the Ehrenfest relation and the effect is almost entirely in the basal plane.

SAMPLE PREPARATION AND CHARACATERIZATION

The sample for this study was made in the same way as described in our previous paper.[2]  Because the isotope $^{10}B$, which is the most abundant isotope in natural boron, has a large neutron absorption cross section, isotopically-enriched $^{11}B$ (Eagle Picher, 98.46 atomic % enrichment) was used for the synthesis.  A mixture of $^{11}B$ powder (less than 200 mesh particle size) and chunks of Mg metal was reacted in a capped BN crucible at 850 C under an argon atmosphere of 50 bar for 1.5 hours. The resulting sample displayed a sharp superconducting transition (0.4 K wide) with an onset at 39 K. Both x-ray and neutron diffraction data showed the sample to be single phase with the hexagonal $AlB_2$-type structure.

NEUTRON POWDER DIFFRACTION AND DATA ANALYSIS

Neutron powder diffraction data were collected as a function of increasing temperature between 3 and 100 K, using the high-resolution diffractometer HRPD at the ISIS facility (Rutherford Appleton Laboratory, UK). Only the highest-resolution back-scattering detector bank was employed (nominal $2\theta=168.33°$), with the sample in "slab" geometry. For this, a thin layer of $Mg^{11}B_2$ powder was compressed between two vanadium foils ($20\times20$ mm$^2$), and completely surrounded by an aluminum frame. A calibrated Fe:Rh thermocouple (accuracy: 0.2 K) was positioned in a hole inside the Al frame, and held in place by conductive Cu-based low-temperature paste. The sample was inserted in an "orange" cryostat, and bathed in He exchange gas. The data collection time was approximately 2 hrs per temperature. Full-pattern Rietveld refinements of the data sets were carried out using the program GSAS.[7] Bragg peaks from the sample were sharp, but somewhat broader than the intrinsic instrument resolution (~0.1% at 1 Å d-spacing). In fact, close inspection of the data reveled that the (00l) reflections were asymmetrically broadened, with a "tail" on the high-d-spacing side. It is unclear whether this results from stacking faults or from strain from grain-interaction stresses, as has been seen in other work on these materials.[8] However, this effect is clearly temperature-independent, and does not affect the determination of the thermal expansion. Therefore, the lattice parameters were determined from single-phase refinements. The statistical precision of the refined lattice parameters is ~1/300000, the absolute accuracy being at least 10 times worse, as is typical for open-geometry instruments.

RESULTS AND DISCUSSION

The refined unit cell volume and hexagonal lattice parameters versus temperature are shown in Figs. 1 and 2. These new results confirm the effects seen in our previous measurement [1] and provide error bars small enough to draw quantitative conclusions. The figures compare the data with two different models for the thermal expansion. In Fig. 1, an Einstein model with a single Einstein frequency of 350 K (as described in Ref. 2) is used to fit the volume thermal expansion at temperatures above $T_c$. For this simple model, near-zero thermal expansion is predicted at low temperatures. The change in thermal expansion beginning at (or perhaps a degree or two above) $T_c$ ($T_c$= 39 K for this sample) is well outside the error bars, resulting in a significant increase in the cell volume and a negative thermal expansion below $T_c$. The dashed line in Fig. 1 shows the change in thermal expansion predicted by the Ehrenfest relation (equation 1). For the calculation, we take $\delta T_c/\delta P$=-1.11 K/GPa [9] and ($C_s$-$C_n$)=133 mJ•mol-1•K-1.[10] Equation (1) defines the expected change in $\delta V/\delta T$ at $T_c$, but does not predict the evolution of the cell volume versus temperature for temperatures well below $T_c$. It is obvious that the effect we observe experimentally is dramatically larger than predicted by the Ehrenfest relation -- perhaps five times larger.

One might ask whether fluctuations near $T_c$ could explain the large effects we see. Fluctuation effects are expected to be important only for materials with a short superconducting coherence length such as the layered copper-oxide superconductors.[4,6] Normally, one would not expect fluctuations to be important for conventional superconductors. However, in a recent study of the specific heat of $MgB_2$, Park et al. [11] concluded that the effects of fluctuations were present extending a few degrees above and below $T_c$. Thus, our observation that changes in the thermal expansion may begin one or two degrees above $T_c$ could be explained by fluctuations. However, in the layered copper-oxide superconductors, fluctuations were observed to increase the change in thermal expansion below $T_c$ by less than a factor of two [9], not a factor of five, as we observe for $MgB_2$. Thus, we consider it unlikely that fluctuations can explain the large change in thermal expansion that we observe.

It is important to ask whether the excess negative thermal expansion below about 40 K might be explained by conventional models not associated with the onset of superconductivity. Fig. 2 compares a simple thermal expansion model based on two phonon frequencies with the experimental results. Negative thermal expansion at low temperature can be explained in terms of a low-energy phonon with a negative Grüneisen parameter. In such a scenario, at higher temperatures, the thermal expansion is positive due to the contribution of many phonons (with positive Grüneisen parameters) while, at low temperatures, the low-energy phonon with negative Grüneisen parameter dominates. An Einstein equation with two phonons is a simple way to model such behavior:

$$\ln\left(\frac{V}{V_0}\right) = \frac{k_B n}{B V_m}\left(\frac{\gamma_1 \theta_1}{e^{(\theta_1/T)}-1} + \frac{\gamma_2 \theta_2}{e^{(\theta_2/T)}-1}\right) \tag{2}$$

and corresponding equations for the lattice parameters, where $V$ is the unit cell volume, $V_0$ is the unit cell volume at T=0, $k_B$ is the Boltzman constant, $n$ is the number of atoms per unit cell, $B$ is the bulk modulus, $V_m$ is the molar volume, and $\gamma_i$ and $\theta_i$ are the Grüneisen parameter and Einstein temperature for the $i$th phonon. For MgB$_2$ we take $V_m$=1.74x10$^{-5}$ m$^3 \cdot$mol$^{-2}$ and $B$=147.2 GPa (based on Ref. 2). Negative thermal expansion at low temperature is given by equation (2) when $\theta_1 > \theta_2$, $\gamma_1 > 0$, and $\gamma_2 < 0$.

The solid lines in Fig. 2 are least-squares fits to the data performed in the following way: Equation (2) is fit to the data for the lattice parameter $a$ versus T, where the largest effect is observed (Fig. 2b) with $a_0$, $\gamma_1$, $\theta_1$, $\gamma_2$, and $\theta_2$ as variables. This gives the two phonon frequencies $\theta_1$=222(70) K and $\theta_2$=69(48) K. These phonon frequencies are then held constant and equation (2) is least-squares fit to the data for the unit cell volume (Fig. 2a) and $c$-axis lattice parameter (Fig 2c) versus temperature. From the fit to the unit cell volume data, one obtains the two Grüneisen parameters, $\gamma_1$=1.33(7) and $\gamma_2$=-0.304(2). (Note that these two phonon frequencies and Grüneisen parameters should be viewed as weighted mean values at high and low temperature, not parameters for actual phonons.) These values are not unreasonable. It is clear that this simple model is adequate to fit the data within the statistical errors; i.e., the data are not of high enough precision to justify a more sophisticated model. Thus, one cannot rule out that most of the change in thermal expansion at low temperatures in MgB$_2$ is simply due to phonon effects and that its onset so close to $T_c$ is accidental.

Negative thermal expansion in the basal plane is, in fact, rather common for layered materials with hexagonal crystal structures.[12] Examples that have been studied for many years include graphite,[13] hexagonal boron nitride,[14] cadmium,[15,16] and zinc.[15,17,18] Interestingly, the first two of these have hexagonal networks of atoms identical to the boron layers in MgB$_2$ (albeit, without the Mg atoms separating the layers and with a different stacking sequence). Abdullaev [19] has recently presented a model to explain the common occurrence of negative thermal expansion in the basal plane in hexagonal layered structures. He provides a simple physical model for this behavior in which he views the hexagonal layer as a membrane. As the membrane is subjected to perpendicular tension it has a tendency to shrink. Abdullaev argues that large transverse thermal vibrations can have the same effect in a layered crystal system. This can give rise to a negative weighted mean Grüneisen parameter in the basal plane. The negative thermal expansion in the basal plane of graphite is explained in this way. Also, ab initio calculations of the lattice dynamics of hexagonal boron nitride have confirmed the existence of the negative Grüneisen parameter for the transverse optic modes.[20] Surprisingly, however, in analytical models for the thermal expansion parallel and perpendicular to the basal plane, Abdullaev shows that a negative mean Grüneisen parameter is not a requirement for negative thermal expansion. In general, for layered crystals with axial symmetry (specifically in the hexagonal crystal system), the thermal expansions along the $a$ and $c$ axes, $\alpha_a$ and $\alpha_c$, can be expressed in the form [19]:

$$\alpha_a = \frac{C_V}{V}\left[\gamma_a f_1(C_{ij}) - \gamma_c f_2(C_{ij})\right]$$
$$\alpha_c = \frac{C_V}{V}\left[\gamma_c f_3(C_{ij}) - \gamma_a f_4(C_{ij})\right] \tag{3}$$

where $C_V$ is the heat capacity, $V$ is the cell volume, $\gamma_a$ and $\gamma_c$ are the weighted mean Grüneisen parameters along the $a$ and $c$ axes, respectively, and $f_n(C_{ij})$ are functions of the elastic constants. Such a model can explain the negative thermal expansion in the basal plane in cadmium and zinc, even though the weighted mean Grüneisen parameters are positive at all temperatures.[21] A key feature of equation (3) is the possibility of a crossover from positive to negative thermal expansion at low temperature. This can occur in two ways: The weighted mean Grüneisen parameter can change sign at low temperature as higher-energy phonons no longer contribute or the subtle balance between the two terms in equation (3) can change. Abdullaev argues that the latter situation occurs in cadmium, where the basal-plane thermal expansion crosses over from positive to negative at about 40 K.[12,15,16]

CONCLUSIONS

Our experimental results show that $MgB_2$ exhibits a negative thermal expansion in the basal plane, whose onset is very near $T_c$, but whose magnitude is much larger than predicted by the Ehrenfest relation. We conclude that the change in thermal expansion required by the Ehrenfest relation, which must be present, is overwhelmed by an additional negative contribution that is approximately five times larger. If $MgB_2$ were not a superconductor, one might conclude that its negative thermal expansion at low temperature comes simply from phonon behavior, as is observed in many other hexagonal layered crystal systems. The crossover from positive to negative thermal expansion in the basal plane at about 40 K is not unprecedented for such systems and could have a simple explanation that is not connected with the onset of superconductivity. Model calculations could be done if values were available for the weighted mean Grüneisen parameters in the layers and normal to the layers and the elastic constants, including off-diagonal terms. However, two observations suggest that the anomaly in thermal expansion could be related to the superconducting transition. First, in the raw data, the onset of the change in thermal expansion occurs very near $T_c$ -- perhaps, at most, one or two degrees above $T_c$. The small discrepancy in temperature could be explained by fluctuations. Second, the strong anisotropy of the thermal-expansion anomaly, with the effect being dramatically larger in the basal plane (along the *a* axis), agrees qualitatively with what is known about superconductivity in $MgB_2$. Superconductivity is believed to result from coupling between electrons in the boron σ bands and the $E_{2g}$ boron in-plane breathing-mode phonons.[22] The onset of superconductivity, thus, removes σ-bonding electrons, weakening the B-B bonds and resulting in expansion in the basal plane. The $E_{2g}$ phonons are known to be very anharmonic.[23] Thus, the flat-bottomed potential gives a very small restoring force for small displacements and an unusually large effect would be expected. Perhaps, it is possible to reconcile the connection between the onset of superconductivity and the crossover to negative thermal expansion in the basal plane. As Abdullaev has shown,[19] such a crossover can occur when temperature-dependent changes in the phonon spectrum modify the subtle balance of terms that define the thermal expansion. It is not impossible that the onset of superconductivity triggers this crossover in $MgB_2$ because it is a compound that is already on the verge of manifesting such behavior. Careful measurements of the low-temperature thermal expansion of other superconducting and non superconducting hexagonal diboride compounds could shed additional light on this question.

ACKNOWLEDGEMENTS

This work was supported by the US Department of Energy, Office of Basic Energy Sciences, contract No. W-31-109-ENG-38.

FIGURE CAPTIONS

Fig. 1.  Cell volume of $MgB_2$ versus temperature as determined from Reitveld refinement using high-resolution neutron powder diffraction data.  The solid curve is a least-squares fit using an Einstein model with a single phonon energy.  The dashed curve shows the predicted change in volume thermal expansion at $T_c$ based on the Ehrenfest relation.

Fig. 2.  (a) Cell volume, $V$, and lattice parameters (b) *a* and (c) *c* of $MgB_2$ versus temperature.  The solid curves are Einstein models with two phonon energies where the Grüneisen parameter is positive for the phonon with the higher energy and negative for the phonon with the lower energy.

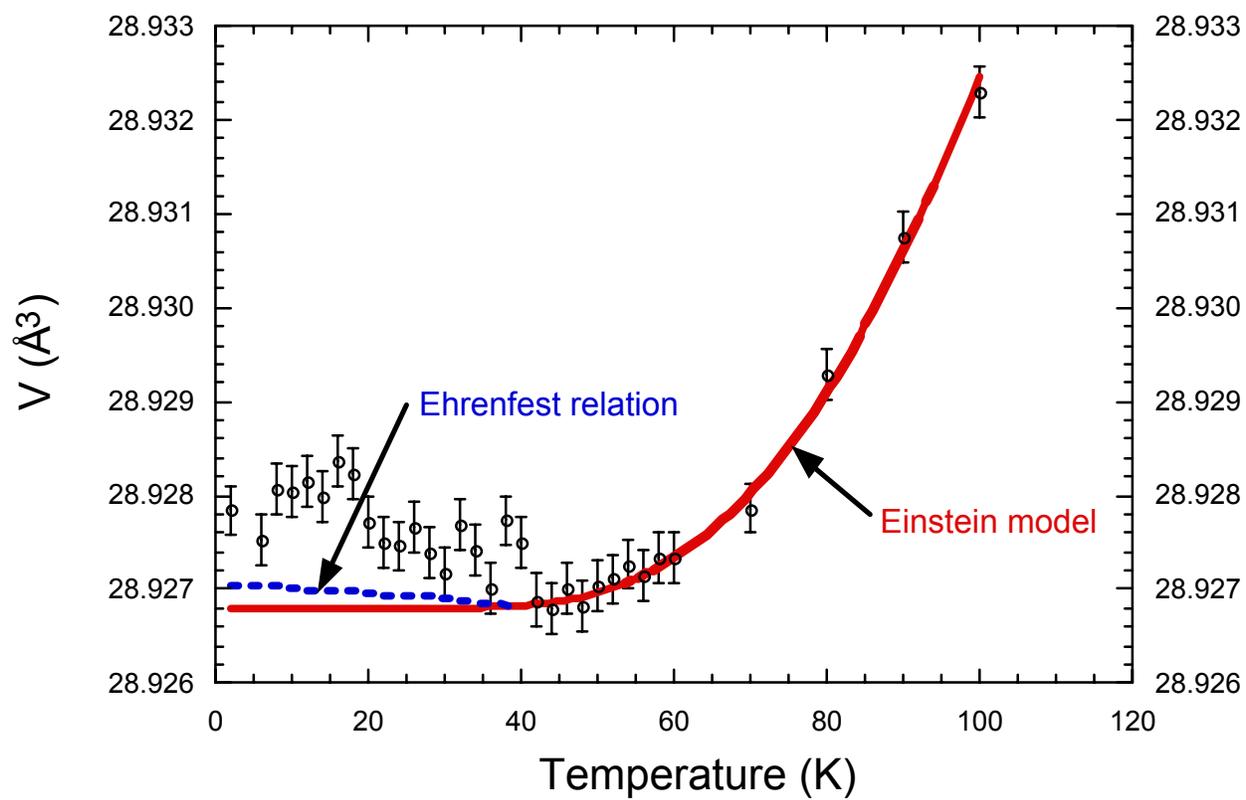

Fig. 1.

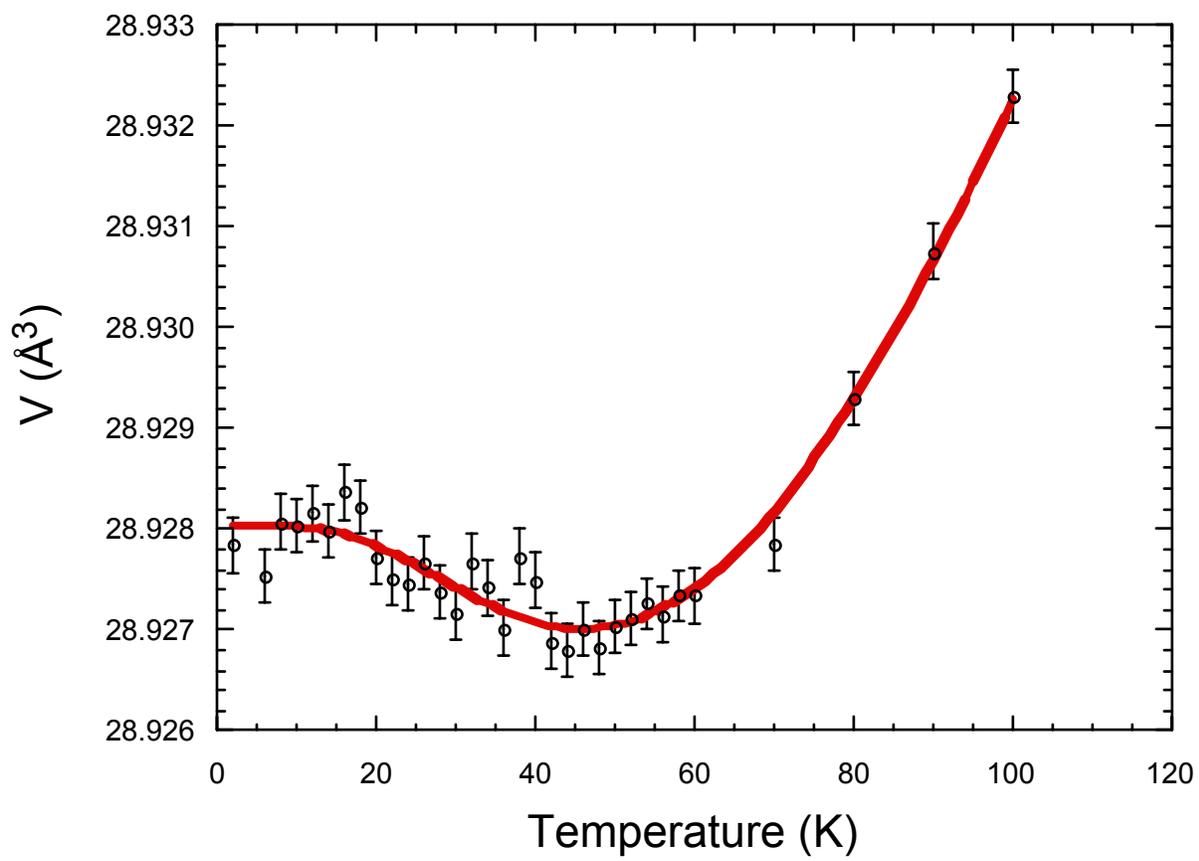

Fig. 2a.

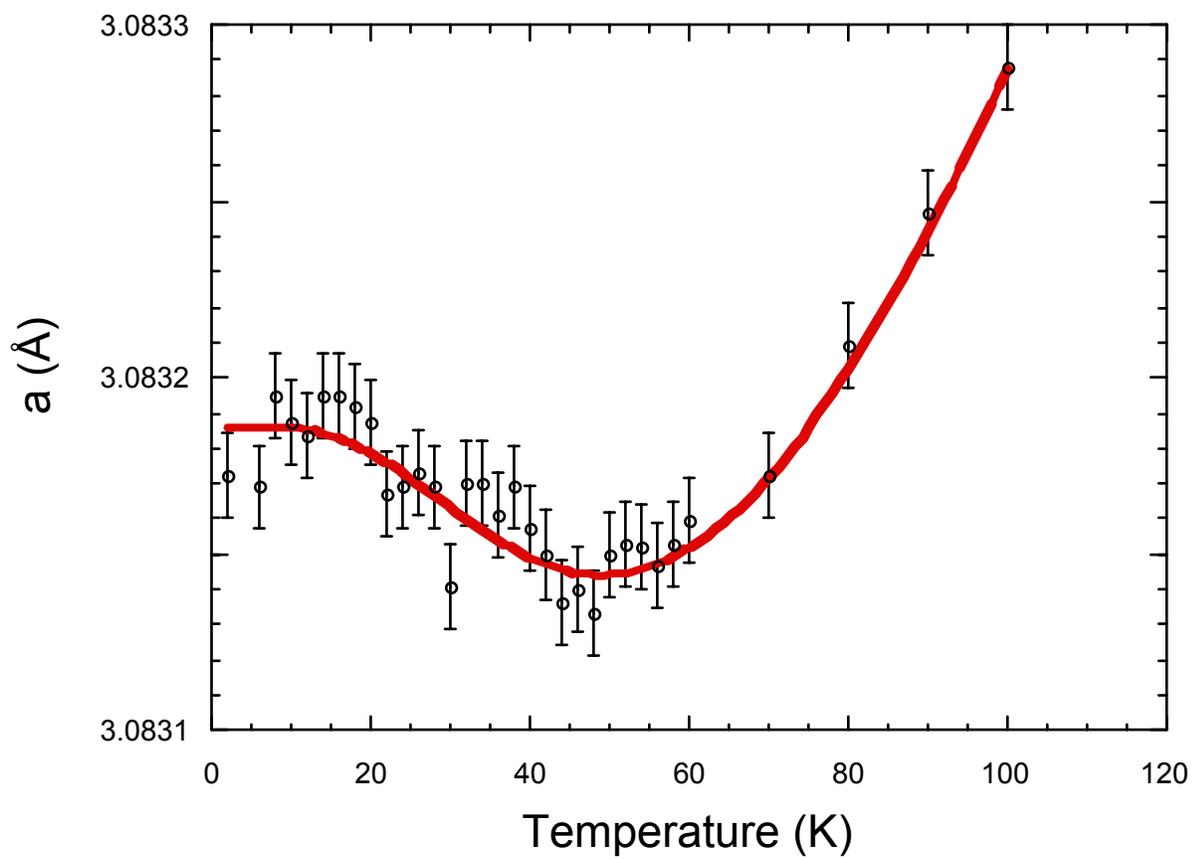

Fig. 2b.

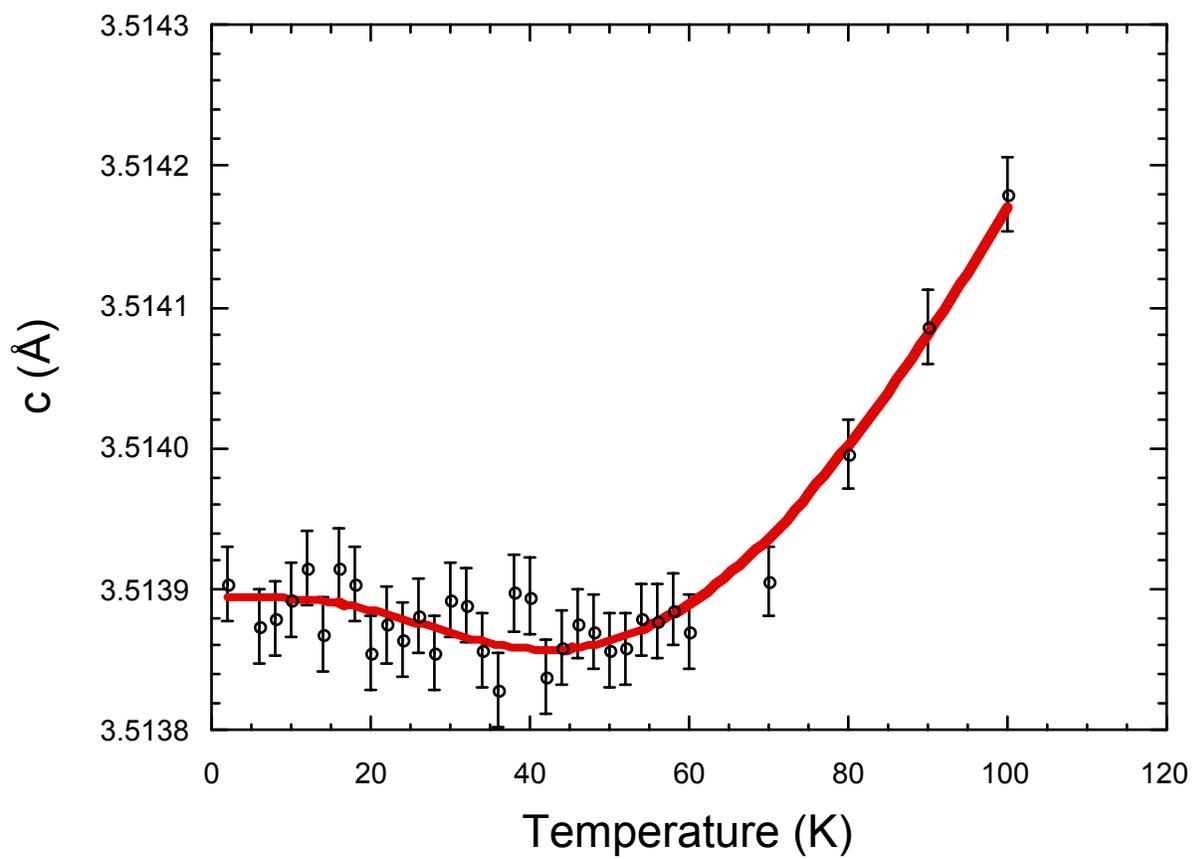

Fig. 2c.